\documentclass{ifacconf}

\usepackage{graphicx}      
\usepackage{natbib}        
\usepackage{epstopdf}
\usepackage{epsfig} 
\usepackage{tikz}
\usepackage{subcaption}
\usepackage{psfrag}
\usepackage[width=.49\textwidth]{caption}
\usepackage{lipsum}
\usepackage{url}
\usepackage{amsmath,amsfonts,amssymb,color}

\newtheorem{theorem}{Theorem}
\newtheorem{corollary}{Corollary}
\newtheorem{lemma}{Lemma}
\newtheorem{remark}{Remark}
\newtheorem{definition}{Definition}

\newtheorem{proposition}{Proposition}
\newtheorem{assumption}{Assumption}
\usepackage{tikz}
\usetikzlibrary{positioning}
\usetikzlibrary{arrows}

\newcommand{\ignore}[1]{}

\DeclareMathOperator*{\argmin}{argmin}

\renewcommand{\qed}{\hfill\ensuremath{\blacksquare}}

\allowdisplaybreaks
\begin{document}
\begin{frontmatter}

\title{Game-Theoretic Choice of Curing Rates Against Networked SIS Epidemics by Human Decision-Makers} 

\thanks[footnoteinfo]{This research was supported in part by the National Science Foundation, under grant CNS-1718637. This is an extended version of a paper that appears in the proceedings of the 2nd IFAC Conference on Cyber-Physical \& Human Systems, 2018.}

\author[First]{Ashish R. Hota}
\author[Second]{Shreyas Sundaram} 

\address[First]{Automatic Control Laboratory, ETH Zurich, Switzerland (Email: ahota@control.ee.ethz.ch).}
\address[Second]{School of Electrical and Computer Engineering, Purdue University, USA (Email: sundara2@purdue.edu)}

\begin{abstract}                
We study networks of human decision-makers who independently decide how to protect themselves against Susceptible-Infected-Susceptible (SIS) epidemics. Motivated by studies in behavioral economics showing that humans perceive probabilities in a nonlinear fashion, we examine the impacts of such misperceptions on the equilibrium protection strategies. In our setting, nodes choose their curing rates to minimize the infection probability under the degree-based mean-field approximation of the SIS epidemic plus the cost of their selected curing rate. We establish the existence of a degree based equilibrium under both true and nonlinear perceptions of infection probabilities (under suitable assumptions). When the per-unit cost of curing rate is sufficiently high, we show that true expectation minimizers choose the curing rate to be zero at the equilibrium, while curing rate is nonzero under nonlinear probability weighting.
\end{abstract}


\end{frontmatter}

\section{Introduction}
\label{section:introduction}
Factors that influence the security, robustness and resilience of networked socio-cyber-physical systems include the characteristics of threats and attacks \citep{pastor2015epidemic,la2016interdependent}, topology of the network \citep{hota2016interdependent,drakopoulos2016network}, and centralized vs. decentralized decision-making \citep{manshaei2013game}. In addition, decisions made by humans that interact and use these systems also have a significant impact on their security and resilience \citep{hota2017impacts,sanjab2017prospect}. In this paper, we investigate the impacts of human decision-making in the context of Susceptible-Infected-Susceptible (SIS) epidemics.

SIS epidemics capture a wide range of dynamics in cyber-physical and social networks, such as spread of diseases in human society \citep{hethcote2000mathematics}, and viruses in computer networks \citep{sellke2008modeling}. There is a large literature on mean-field approximations, characterizations of steady-state behavior, and centralized protection strategies to control SIS epidemics \citep{preciado2014optimal,pastor2001epidemic,van2009virus,khanafer2016stability}; see \citep{nowzari2016analysis,pastor2015epidemic} for recent reviews. 

While centralized protection strategies may not be practical for large-scale networked systems, decentralized and game-theoretic protection strategies against network epidemics have been relatively less explored \citep{nowzari2016analysis,pastor2015epidemic}. A common assumption in the existing literature is that the decision-makers are risk neutral (i.e., expected cost minimizers), and perceive infection probabilities as their true values. However, there is a large body of work in psychology and behavioral economics that has shown that humans perceive probabilities differently from their true values \citep{kahneman1979prospect,dhami2016foundations,barberis2013thirty} (see Section \ref{section:probweighting} for further details), and these behavioral aspects of decision-making often have a significant impact on the security of networked systems  \citep{hota2016interdependent,hota2016fragility}. In the context of epidemics, there is a related body of research that investigates certain human aspects of decision-making, particularly imitation behavior \citep{mbah2012impact}, and empathy \citep{eksin2016disease} in an (evolutionary) game-theoretic framework. On the other hand, the impacts of human (mis)-perception of probabilities is little explored in the existing work.

Our goal, in this paper, is to characterize the impacts of human perception of infection probabilities (captured by prospect-theoretic probability weighting functions \citep{kahneman1979prospect}) on their protection strategies against SIS epidemics on networks, and compare it with the equilibria without probability weighting. Under SIS epidemics, each node in the network can be in one of the two states, i) susceptible, and ii) infected. An infected node is cured with a {\it curing rate} $\delta \geq 0$, while a susceptible node becomes infected following a Poisson process with rate $\nu$ per infected neighbor. We consider a protection strategy where nodes choose their curing rates strategically.\footnote{In \cite{hota2018game}, we considered the setting where nodes choose whether or not to vaccinate against SIS epidemics.} Since we consider a cost minimization problem for the decision-makers, we refer to players who perceive probabilities as their true values as {\it true expectation minimizers}. 

Prior work \citep{omic2009protecting,trajanovski2015decentralized} on epidemic games has relied on the N-Intertwined Mean Field Approximation (NIMFA) \citep{van2009virus,van2013homogeneous}. \citep{omic2009protecting} studied a game-theoretic setting where nodes choose their curing rates, and showed the existence of a pure Nash equilibrium (PNE) assuming that the steady-state infection probability of a node is a convex function of her own curing rate under the NIMFA. However, the follow up work \citep{van2013homogeneous} observed that the above convexity assumption does not hold in general. Furthermore, under the NIMFA, the nodes need to be aware of the structure of the entire network.

In order to analyze the game-theoretic setting in general networks and under prospect-theoretic perception of probabilities, we consider the degree-based mean-field (DBMF) approximation \citep{pastor2001epidemic,pastor2015epidemic} (summarized in Section \ref{sub:dbmf}) to capture the infection probabilities. Under the DBMF approximation, each node is only aware of its own degree and the degree distribution of the network. While the DBMF approximation is coarser than the NIMFA, it is more tractable to analyze. In particular, we show that the (perceived) infection probability of a node is convex in her curing rate under the DBMF approximation under suitable assumptions. We then prove the existence of a degree based equilibrium (DBE) (formally defined in Section \ref{section:curing}) for both true expectation minimizers and under nonlinear probability weighting, and derive various characteristics of the DBE. For instance, when the per-unit cost of curing rate is high, true expectation minimizers choose the curing rate to be $0$ at the DBE, while under nonlinear perception of probabilities, the equilibrium curing rate is always nonzero for any finite per-unit cost of curing rate. We further illustrate how the optimal curing rate varies as a function of the cost parameter and the nonlinear probability weighting function in degree-regular graphs.

\section{Preliminaries}
\label{section:prelim}
\subsection{Nonlinear probability weighting}
\label{section:probweighting}
Decades of research in behavioral economics has shown that humans {\it perceive} probabilities associated with uncertain outcomes in a nonlinear fashion \citep{kahneman1979prospect,gonzalez1999shape,dhami2016foundations}. Specifically, humans overweight probabilities that are close to $0$ (referred to as {\it possibility effect}), and underweight probabilities that are close to $1$ (referred to as {\it certainty effect}). In the Prospect theory framework of \citep{kahneman1979prospect}, the authors captured the transformation of true probabilities into {\it perceived probabilities} by an inverse S-shaped probability weighting function $w :[0,1] \to [0,1]$ (i.e., a true probability $x$ is perceived as $w(x)$). Several parametric forms of weighting functions have been proposed in \citep{tversky1992advances,prelec1998probability,gonzalez1999shape}. These weighting functions have the same general shape, and satisfy the following properties \citep{hota2018game}. 

\begin{figure}[t]
	\begin{center}
	\includegraphics[scale=0.65]{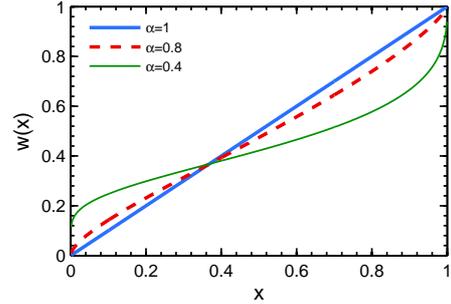}
	\caption{Shape of the probability weighting function \eqref{eq:prelec}. The quantity $x$ is the true probability, and $w(x)$ is the corresponding perceived probability.}\label{fig:prelecweighting}
	\end{center}
\end{figure}

\begin{assumption}\label{assumption:weightingfunction}
The probability weighting function $w$ satisfies the following properties. 
\begin{enumerate}
\item[1.] $w$ is strictly increasing, with $w(0)=0$ and $w(1)=1$.
\item[2.] \sloppy $w'(x)$ has a unique minimum denoted by $x_{\min,w} := \argmin_{x \in [0,1]} w'(x)$. Furthermore, $w'(x_{\min,w}) < 1$ and $w''(x_{\min,w}) = 0$.
\item[3.] $w(x)$ is strictly concave for $x \in [0,x_{\min,w})$, and is strictly convex for $x \in (x_{\min,w},1]$.
\item[4.] $w'(\epsilon) \to \infty$ as $\epsilon \to 0$, and $w'(1-\epsilon) \to \infty$ as $\epsilon \to 0$.
\end{enumerate}
\end{assumption}
The above assumptions imply that there exists a unique $x_{0,w} \in [0,1]$ such that $w(x) > x$ for $x\in[0,x_{0,w})$, and $w(x) < x$ for $x\in(x_{0,w},1]$.

Our theoretical results hold for probability weighting functions that satisfy Assumption \ref{assumption:weightingfunction}. For instance, the weighting function proposed by \citep{prelec1998probability} is given by
\begin{equation}
w(x) = \exp(-(-\ln(x))^\alpha), \text{\qquad} x \in [0,1],
\label{eq:prelec}
\end{equation}
where $\alpha \in (0,1)$, and $\exp(\cdot)$ is the exponential function. For $\alpha=1$, we have $w(x) = x$, i.e., the perceived and true probabilities coincide. For smaller $\alpha$, the function $w(x)$ has a sharper overweighting of low probabilities and underweighting of high probabilities. Figure \ref{fig:prelecweighting} shows the shape of the Prelec weighting function for different values of $\alpha$. Prelec weighting functions with $\alpha \in (0,1)$ satisfy Assumption \ref{assumption:weightingfunction}, and have $x_{\min,w} = x_{0,w} = \frac{1}{e}$, $w(\frac{1}{e}) = \frac{1}{e}$ for every $\alpha \in (0,1)$, and $w'(\frac{1}{e}) = \alpha$. 

\ignore{\begin{assumption}\label{assumption:weightingfunction}
The probability weighting function $w(x)$ satisfies the following properties. 
\begin{enumerate}
\item[1.] \sloppy $w'(x)$ has a unique minimum for $x \in (0,1)$ denoted as $\mathbf{x}_{\min,w} := \argmin_{x \in [0,1]} w'(x)$, and $w''(\mathbf{x}_{\min,w}) = 0$. Furthermore, $w'(\mathbf{x}_{\min,w}) < 1$.
\item[2.] $w(x)$ is strictly concave for $x \in [0,\mathbf{x}_{\min,w})$, and is strictly convex for $x \in (\mathbf{x}_{\min,w},1]$.
\item[3.] $w'(\epsilon) \to \infty$ as $\epsilon \to 0$, and $w'(1-\epsilon) \to \infty$ as $\epsilon \to 0$.
\item[4.] There exists a unique $\mathbf{x}_{0,w} \in [0,1]$ such that $w(x) > x$ for $x\in[0,\mathbf{x}_{0,w})$, and $w(x) < x$ for $x\in(\mathbf{x}_{0,w},1]$.
\end{enumerate}
\end{assumption}}

\subsection{Degree based mean-field approximation of the SIS epidemic}
\label{sub:dbmf}

Consider an undirected network with $\mathcal{D}$ being the set of degrees of the nodes, and degree distribution $P(\cdot)$, i.e., the probability that a randomly chosen node has degree $k$ is $P(k)$. Let $1 < \hat{d} < \infty$ and $\bar{d} < \infty$ be the average and highest degrees of the nodes in the network, respectively. Unless specified otherwise, we assume that the minimum degree of any node in the network is $1$. Furthermore, let the network be {\it uncorrelated}, i.e., the probability that an edge originating from a node with degree $k$ is connected to a node with degree $k'$ is independent of $k$. For uncorrelated networks, the probability that a randomly chosen neighbor (of any node) has degree $i$ is approximately $q_i := \frac{iP(i)}{\langle d\rangle}$ \citep{pastor2015epidemic}, where $\langle d\rangle = \hat{d}-1$.

As discussed earlier, each node in the network can be in one of two states: i) susceptible, or ii) infected. Without loss of generality, let the infection rate to be $\nu = 1$. Under the DBMF approximation \citep{pastor2001epidemic,pastor2015epidemic}, every node with a given degree $k$ is treated as statistically equivalent. Let $\delta_k \geq 0$ be the curing rate of every node with degree $k$. Let $\delta = \{\delta\}_{k \in \mathcal{D}}$ be the vector of curing rates. The infection probability of a degree $k$ node, $\tilde{x}_k(t,\delta)$, evolves as
\begin{align}\label{eq:app_dbmf}
\frac{\partial \tilde{x}_k(t,\delta)}{\partial t} = -\delta_k \tilde{x}_k(t,\delta) + (1-\tilde{x}_k(t))k \sum_{i \in \mathcal{D}} q_i\tilde{x}_i(t,\delta),
\end{align}
under the DBMF approximation. The DBMF approximation is better if the timescale at which nodes interact with each other in the random graph model is faster than the timescale at which the epidemic spreads \citep{pastor2015epidemic}. At the stationary-state of the above dynamics, the infection probability of a degree $k$ node is
\begin{equation}\label{eq:infec_prob}
x_k(\delta) \!= \frac{kv}{\delta_k + kv}, v(\delta) \!= \sum^{\bar{d}}_{i=1} x_i(\delta) q_i \!= \sum^{\bar{d}}_{i=1} \frac{iv(\delta)q_i}{\delta_i + iv(\delta)}.
\end{equation}
The quantity $v(\delta)$ represents the steady-state probability that a randomly chosen neighbor is infected, and satisfies
\begin{equation}\label{eq:v_consistency}
v(\delta) \left[1-\sum^{\bar{d}}_{i=1} \frac{iq_i}{\delta_i + iv(\delta)} \right] = 0. 
\end{equation}
Note that $v(\delta)=0$ always satisfies the above equation, which corresponds to the disease-free state. Furthermore, depending on $\delta$, there may exist a nonzero $v \in (0,1]$ that satisfies \eqref{eq:v_consistency}. A nonzero solution of $v(\delta)$ is referred to as the ``endemic" state where the epidemic persists in the network for a long time. We state the following result on the uniqueness and stability of the endemic state.

\begin{theorem}\label{theorem:app_bullo}
Let $R := \sum_{i\in\mathcal{D}} \frac{iq_i}{\delta_i}$.  
\begin{enumerate}
\item $x^*_i = 0, \forall i \in \mathcal{D}$ is the unique stationary-state of the dynamics in \eqref{eq:app_dbmf} if and only if $R \leq 1$. This disease free state is globally asymptotically stable.
\item If $R > 1$, $x^*_i = 0, \forall i \in \mathcal{D}$ is an unstable stationary-state. Furthermore, there exists a stationary state, referred to as an endemic state, where $x^{*}_i > 0, \forall i \in \mathcal{D}$ if and only if $R > 1$. This nonzero endemic state is unique, is locally exponentially stable, and the dynamics converge to this endemic state from any initial condition except the disease free state. 
\end{enumerate}
\end{theorem}
The proof exploits the relationship between the NIMFA and DBMF approximations, and leverages similar results obtained for the NIMFA \citep{khanafer2016stability,bullo2016lectures}. We omit this for space constraints as it is analogous to the proof of \cite[Theorem 1]{hota2018game}.

Following conventional notation, we denote the vector of curing rates by all nodes other than the nodes with degree $k$ as ${\delta}_{-k}$. We start with a corollary of Theorem \ref{theorem:app_bullo}. 

\begin{corollary}\label{existence:v_metastable}
Let $\hat{\delta}_k({\delta}_{-k}) := kq_k \left[1 - \sum^{\bar{d}}_{\substack{i = 1 \\ i \neq k}} \frac{iq_i}{\delta_i} \right]^{-1}$. A unique nonzero solution of $v(\delta_k,{\delta}_{-k})$ to \eqref{eq:v_consistency} exists if and only if $\delta_k \in [0,\hat{\delta}_k({\delta}_{-k}))$. 
\end{corollary}
\begin{pf}
Note that $\delta_k < \hat{\delta}_k({\delta}_{-k})$ is equivalent to
\begin{align*}
\frac{\delta_k}{kq_k} < \left[1 - \sum^{\bar{d}}_{\substack{i = 1 \\ i \neq k}} \frac{iq_i}{\delta_i} \right]^{-1} \iff & \frac{kq_k}{\delta_k} + \sum^{\bar{d}}_{\substack{i = 1 \\ i \neq k}} \frac{iq_i}{\delta_i} > 1,
\end{align*}
or $R > 1$. Following Theorem \ref{theorem:app_bullo}, there exists a unique endemic state corresponding to a unique nonzero $v(\delta)$. 
\qed \end{pf}

We now show monotonicity and convexity of $v(\delta)$ in the endemic state. We denote $\frac{\partial v}{\partial \delta_k}$ by $v'_k$ and $\frac{\partial^2 v}{\partial \delta^2_k}$ by $v''_k$.  

\begin{lemma}\label{lemma:neighbor_prob_convexity}
$v(\delta_k,\delta_{-k})$ is decreasing and convex in $\delta_k$ for $\delta_k \in [0,\hat{\delta}_k(\delta_{-k}))$. 
\end{lemma}
\begin{pf}
We drop the argument $(\delta_k,\delta_{-k})$ from the proof for better readability. From \eqref{eq:v_consistency}, we know that a nonzero $v(\delta)$ must satisfy
\begin{align}
& 1 = \sum^{\bar{d}}_{i=1} \frac{iq_i}{\delta_i + iv(\delta)} \label{eq:consistency}
\\ \implies & 0 = \sum^{\bar{d}}_{\substack{i = 1 \\ i \neq k}} \left[ - \frac{i^2 v'_k q_i}{(\delta_i + iv)^2} \right] - \frac{k (1+kv'_k) q_k}{(\delta_k+ kv)^2} \nonumber
\\ \implies & \frac{k (1+kv'_k) q_k}{(\delta_k+ kv)^2} = - \sum^{\bar{d}}_{\substack{i = 1 \\ i \neq k}} \frac{i^2 v'_k q_i}{(\delta_i + iv)^2} \label{eq:kvkp}
\\ \implies & -v'_k =  \frac{k q_k}{(\delta_k+ kv)^2} \left[ \sum^{\bar{d}}_{i=1}  \frac{i^2 q_i}{(\delta_i + iv)^2} \right]^{-1} > 0. \label{eq:vkp}
\end{align}
We then differentiate \eqref{eq:kvkp} with respect to $\delta_k$, and obtain
\begin{align*}
& \frac{k^2q_k v''_k}{(\delta_k+kv)^2} - \frac{2kq_k (1+kv'_k)^2}{(\delta_k+kv)^3} 
\\ & \qquad = \sum^{\bar{d}}_{\substack{i = 1 \\ i \neq k}} \left[ - \frac{i^2 q_i v''_k}{(\delta_i + iv)^2} + \frac{2i^3q_i(v'_k)^2}{(\delta_i+iv)^3} \right]
\\ \implies & v''_k \sum^{\bar{d}}_{i=1} \frac{i^2 q_i}{(\delta_i + iv)^2} = \frac{2kq_k (1+kv'_k)^2}{(\delta_k+kv)^3} + \sum^{\bar{d}}_{\substack{i = 1 \\ i \neq k}} \frac{2i^3q_i(v'_k)^2}{(\delta_i+iv)^3},
\end{align*}
following straightforward calculations.Thus, $v''_k > 0$. 
\qed
\end{pf}

\begin{remark}\label{remark:nonzero_v}
In the rest of this paper, we define $v(\delta)$ as the nonzero solution that satisfies \eqref{eq:v_consistency} if $1 < \sum^{\bar{d}}_{i=1} \frac{iq_i}{\delta_i}$, and $v(\delta) = 0$ otherwise. In other words, $v(\delta) := \max(0,\hat{v}(\delta))$, where $z = \hat{v}(\delta) \in \mathbb{R}$ is the unique root of $1 - \sum^{\bar{d}}_{i=1} \frac{iq_i}{\delta_i + iz} = 0$. Accordingly, both $\hat{v}(\delta)$ and $v(\delta)$ are continuous in $\delta$.
\end{remark}

\section{Strategic Choice of Curing Rate}
\label{section:curing}

\subsection{Equilibria without probability weighting}

Let $\mathcal{D} \subseteq \{1,2,\ldots,{\bar{d}}\}$ with ${\bar{d}} < \infty$ be the set of degrees of the network. We assume that each node is only aware of her own degree, and the degree distribution $P(\cdot)$. Therefore, all nodes with a given degree have the same information about the network. This is more realistic assumption in large-scale systems compared to assuming that all nodes know the entire network topology (which is the case in related prior work on epidemic games \citep{omic2009protecting}). We assume that all nodes with degree $k$ choose a curing rate $\delta_k \geq 0$ as a pure strategy, i.e., they behave as if being controlled by a single entity. Accordingly, under the DBMF approximation, all degree $k$ nodes experience an identical infection probability in the endemic state. 

Let $c_k > 0$ be the per-unit cost of curing rate for nodes with degree $k$. In this subsection, to establish a baseline, we consider nodes who minimize the infection probability in the endemic state plus the cost of their selected curing rate, i.e., they are true expectation minimizers. We will later compare this to the outcome under nonlinear probability weighting. The expected cost of nodes with degree $k$ is defined as
\begin{equation}\label{eq:cost_fun}
J_k(\delta_k,\delta_{-k}) := x_k(\delta_k,\delta_{-k}) + c_k \delta_k,
\end{equation}
where $x_k(\delta_k,\delta_{-k})$ is the steady-state infection probability of degree $k$ nodes as defined in \eqref{eq:infec_prob}. Note that when $\delta_k = 0$, $x_k(0,\delta_{-k}) = 1$, and $J_k(0,\delta_{-k}) = 1$. Consequently, it is never optimal to choose $\delta_k > \frac{1}{c_k}$. Therefore, we define the set of feasible curing rates $\delta_k$ as $\Delta_k := [0,\frac{1}{c_k}]$. Furthermore, we assume that the nodes prefer to choose $\delta_k = 0$ instead of $\frac{1}{c_k}$ when the optimal cost is $1$.
 
\sloppy
We denote the game defined above by $\Gamma(\mathcal{D},P,\{c_k\}_{k \in \mathcal{D}})$. We now define the degree based equilibrium (DBE) of this game in a manner analogous to the definition of a pure Nash equilibrium (PNE) for strategic games.
\begin{definition}
The vector of curing rates $\delta^{\mathtt{NE}}$, with $\delta_k^{\mathtt{NE}} \in \Delta_k $, is a DBE if $J_k(\delta_k^{\mathtt{NE}},\delta_{-k}^{\mathtt{NE}}) \leq J_k({\delta}_k,\delta_{-k}^{\mathtt{NE}})$ for every ${\delta}_k \in \Delta_k, k \in \mathcal{D}$. Note that all nodes of the same degree choose the same curing rate in a DBE. 
\end{definition}

\begin{remark}\label{remark:dbe}
The above definition differs from the standard notion of PNE, where each node can potentially choose a different strategy (depending on the choices of the other nodes). Nonetheless, the notion of DBE is {\it mathematically equivalent} to a PNE in a game where a single player chooses the curing rate of all nodes of the same degree in order to minimize \eqref{eq:cost_fun}. 
\end{remark}

We now establish the convexity of $x_k(\delta_k,{\delta}_{-k})$ in $\delta_k$ under the DBMF approximation. 

\begin{lemma}\label{lemma:inf_prob_convexity}
$x_k(\delta_k,\delta_{-k})$ is decreasing and convex in $\delta_k$ for $\delta_k \in [0,\hat{\delta}_k(\delta_{-k}))$. 
\end{lemma}
\begin{pf}
Recall from Corollary \ref{existence:v_metastable} that for $\delta_k \in [0,\hat{\delta}_k(\delta_{-k}))$, $v(\delta_k,\delta_{-k})$ is nonzero. We drop the argument $(\delta_k,\delta_{-k})$ for ease of readability, and differentiate the first equation in \eqref{eq:infec_prob} with respect to $\delta_k$ as
\begin{align*}
\frac{\partial x_k}{\partial \delta_k} & =  \frac{kv'_k}{\delta_k+kv} - \frac{kv(1+kv'_k)}{(\delta_k+kv)^2} = \frac{k (\delta_k v'_k - v)}{(\delta_k+kv)^2}.
\end{align*}
From Lemma~\ref{lemma:neighbor_prob_convexity}, we have $v'_k  < 0$, and accordingly $\frac{\partial x_k}{\partial \delta_k} < 0$. 

We now compute
\begin{align*}
\frac{\partial^2 x_k}{\partial \delta^2_k} & = \frac{k \delta_k v''_k}{(\delta_k+kv)^2} - \frac{2k (\delta_k v'_k - v)(1+kv'_k)}{(\delta_k+kv)^3}.
\end{align*}
Note that $\delta_k v'_k - v < 0$ from the above discussion. From Lemma~\ref{lemma:neighbor_prob_convexity}, we have $v''_k  > 0$, and $(1+kv'_k)> 0$ (from \eqref{eq:kvkp} in the proof of Lemma~\ref{lemma:neighbor_prob_convexity}). Accordingly, $\frac{\partial^2 x_k}{\partial \delta^2_k} > 0$.
\qed \end{pf}

With the above result, we now establish the existence of a DBE of the game $\Gamma(\mathcal{D},P,\{c_k\}_{k \in \mathcal{D}})$. 

\begin{proposition}\label{proposition:NEexistence_riskneutral}
$\Gamma(\mathcal{D},P,\{c_k\}_{k \in \mathcal{D}})$ possesses a DBE.
\end{proposition}
\begin{pf}
Consider the set of nodes with degree $k \in \mathcal{D}$. The corresponding feasible strategy set $\Delta_k$ is compact and convex. Following Remark \ref{remark:nonzero_v}, $x_k(\delta)$, and therefore $J_k(\delta)$, is continuous in $\delta \in \prod_{i \in \mathcal{D}} \Delta_i$. 

For a given $\delta_{-k}$, let $\hat{\delta}_k(\delta_{-k})$ be as defined in Corollary \ref{existence:v_metastable}. From Lemma~\ref{lemma:inf_prob_convexity}, it follows that $x_k(\delta_k,\delta_{-k})$, defined as \eqref{eq:infec_prob}, is nonzero, continuous, strictly decreasing and convex in $\delta_k$ for $\delta_k \in [0,\hat{\delta}_k(\delta_{-k}))$. If $\hat{\delta}_k(\delta_{-k}) > \frac{1}{c_k}$, then $x_k(\delta_k,\delta_{-k})$ is convex for $\delta_k \in \Delta_k$. 

On the other hand, suppose $\hat{\delta}_k(\delta_{-k}) \leq \frac{1}{c_k}$. Then, $x_k(\delta_k,\delta_{-k})$ is a continuous and convex function; it is nonzero and convex for $\delta_k \in [0,\hat{\delta}_k(\delta_{-k}))$ (Lemma~\ref{lemma:inf_prob_convexity}), and $x_k(\delta_k,\delta_{-k}) = 0$ for $\delta_k \geq \hat{\delta}_k(\delta_{-k})$ (Corollary~\ref{existence:v_metastable}). Moreover, the derivative of $x_k(\delta_k,\delta_{-k})$ is nondecreasing for $\delta_k \in \Delta_k$, and therefore $x_k(\delta_k,\delta_{-k})$ is convex in $\delta_k$. As a result, for a given $\delta_{-k}$, $J_k(\delta_k,\delta_{-k})$ is convex. 

Recall from Remark \ref{remark:dbe} that DBE is equivalent to the PNE of a strategic game where all nodes with a given degree are controlled by a single player. From the above discussion, this equivalent strategic game is an instance of a {\it concave game}. Following \citep{rosen1965existence}, there exists a PNE of the equivalent game and consequently, a DBE exists.
\qed \end{pf}

In the next result, we obtain several characteristics of the curing rates at a DBE. 

\begin{proposition}\label{proposition:NEproperties_riskneutral}
Let $\delta^{\mathtt{NE}}$ denote the curing rates at a DBE of $\Gamma(\mathcal{D},P,\{c_k\}_{k \in \mathcal{D}})$ with $v^{\mathtt{NE}} > 0$. Then,
\begin{enumerate}
\item If $c_i \geq \frac{1}{i}$ for every $i \in \mathcal{D}$, then $\delta^{\mathtt{NE}}_i = 0$ for every $i \in \mathcal{D}$.
\item If $c_i < \frac{1}{i}$, then $\delta^{\mathtt{NE}}_i > 0$.
\item Let $c_i = c$ for every $i \in \mathcal{D}$. If $\delta^{\mathtt{NE}}_j = 0$ for some $j \in \mathcal{D}$, then $\delta^{\mathtt{NE}}_k = 0$ for all $k \in \mathcal{D}$ with $k > j$.
\end{enumerate}
\end{proposition}

\begin{pf}
For the first part of the proof, let $\mathcal{S}$ be the set of players with positive curing rates. Let $\mathcal{S}^c$ be the complement of $\mathcal{S}$.  Note that when $\delta_j = 0$, the expected cost is $J_j(0,\delta_{-k}) = 1$. Accordingly, for $k \in \mathcal{S}$, we have 
\begin{align}
c_k \delta^{\mathtt{NE}}_k + x^{\mathtt{NE}}_k \leq 1 \implies & c_k \delta^{\mathtt{NE}}_k \leq 1 - \frac{k v^{\mathtt{NE}}}{\delta^{\mathtt{NE}}_k + k v^{\mathtt{NE}}} \nonumber
\\ \implies & c_k \leq \frac{1}{\delta^{\mathtt{NE}}_k + k v^{\mathtt{NE}}}. \label{eq:ci1}
\end{align}
On the other hand, from \eqref{eq:v_consistency} we have
\begin{align}
& 1 = \sum_{k \in \mathcal{S}} \frac{kq_k}{\delta^{\mathtt{NE}}_k + k v^{\mathtt{NE}}} + \sum_{i \in \mathcal{S}^c} \frac{q_i}{v^{\mathtt{NE}}} \nonumber
\\ \implies & 1 - \sum_{i \in \mathcal{S}^c} \frac{q_i}{v^{\mathtt{NE}}} \geq \sum_{k \in \mathcal{S}} c_kkq_k \geq \sum_{k \in \mathcal{S}} q_k \label{eq:epi_neutral_property_ne}
\\ \implies & 1 - \sum_{k \in \mathcal{S}} q_k \geq \frac{1}{v^{\mathtt{NE}}} \sum_{i \in \mathcal{S}^c} q_i \implies v^{\mathtt{NE}} \geq 1, \nonumber
\end{align}
which is true only when $\mathcal{S}$ is an empty set. In \eqref{eq:epi_neutral_property_ne}, the first inequality is a consequence of \eqref{eq:ci1}, and the second inequality is a consequence of $c_k \geq \frac{1}{k}$ and $k \geq 1$.

For the second part of the proof, we compute the derivative of the cost function $J_k(\delta_k,\delta_{-k})$ in \eqref{eq:cost_fun} at $\delta_k = 0$ as
\begin{align*}
\frac{\partial J_k}{\partial \delta_k} = c_k + \frac{\partial x_k}{\partial \delta_k}\bigg\rvert_{\delta_k = 0} = c_k - \frac{1}{kv} < c_k - \frac{1}{k} < 0.
\end{align*}
Therefore, $\delta_k=0$ is not the optimal curing rate irrespective of $\delta_{-k}$. Finally, let $\delta^{\mathtt{NE}}_j = 0$ for a node with degree $j$. Then, $\frac{\partial J_j}{\partial \delta_j}\bigg\rvert_{\delta_j = 0} = c - \frac{1}{jv^{\mathtt{NE}}} \geq 0$. Now, for any $k > j$, we have $c - \frac{1}{kv^{\mathtt{NE}}} > c - \frac{1}{jv^{\mathtt{NE}}} > 0$. Thus, we have $\delta^{\mathtt{NE}}_k = 0$.
\qed \end{pf}

The second property and a weaker version of the first property stated in the above proposition were also shown in \citep{omic2009protecting} under the NIMFA of the SIS dynamics. Proposition \ref{proposition:NEproperties_riskneutral} shows that these properties also hold under the DBMF approximation. 

The third part of the above result shows that when all nodes have homogeneous per-unit curing costs, and the equilibrium curing rate is $0$ for certain degrees of nodes, then these nodes must correspond to a set of high degree nodes. {\bf Intuitively, for nodes with a large number of neighbors, increasing their curing rates has limited impact on counteracting the relatively high probability of infection they are exposed to via their neighbors.} 

\subsection{Equilibria under probability weighting}

In this subsection, we establish the existence of a DBE when the nodes have nonlinear perception of infection probabilities. As discussed in Section \ref{section:probweighting}, we consider probability weighting functions that satisfy Assumption \ref{assumption:weightingfunction}. Let the weighting function for the set of nodes with degree $k$ be $w_k(\cdot)$. Let $c_k > 0$ denote the per-unit cost of curing rate as before. The perceived expected cost incurred by this set of nodes is defined as
\begin{equation}\label{eq:cost_prelec}
J^{(w)}_k(\delta_k,\delta_{-k}) := w_k(x_k(\delta_k,\delta_{-k})) + c_k \delta_k.
\end{equation}
The set of feasible curing rates $\delta_k$ is $\Delta_k := [0,\frac{1}{c_k}]$. We denote the resulting game as $\Gamma(\mathcal{D},P,\{c_k\}_{k \in \mathcal{D}},\{w_k\}_{k \in \mathcal{D}})$.

Recall from Assumption \ref{assumption:weightingfunction} that $w_k(x)$ is concave for $x \in [0,\mathbf{x}_{\min,w_k}]$ and is convex for $x \in [\mathbf{x}_{\min,w_k},1]$, where $\mathbf{x}_{\min,w_k} := \argmin_{x \in [0,1]} w'_k(x)$. Therefore, the cost function in \eqref{eq:cost_prelec} is not necessarily convex for $\delta_k \geq 0$, unlike the cost function for true expectation minimizers. In order to establish the existence of a DBE under nonlinear probability weighting, we start with the following proposition. 

\begin{proposition}\label{proposition:restrict_c}
For a given $z \in (0,1)$, let $c_k = c_0 > \frac{1}{(1-z)}$ for every $k \in \mathcal{D}$. Then, for every $\delta \in \prod_{i \in \mathcal{D}} \left[0,\frac{1}{c_0}\right]$ and $k \in \mathcal{D}$, $x_k(\delta) > z$.  
\end{proposition}
\begin{pf}
Let $\delta_0$ be the vector of curing rates with $\delta_i = \frac{1}{c_0}$, $\forall i \in \mathcal{D}$. For $\delta \in \prod_{i \in \mathcal{D}} \left[0,\frac{1}{c_0}\right]$, we have $v(\delta_0) \leq v(\delta)$ (Lemma \ref{lemma:neighbor_prob_convexity}), and thus, $x_k(\delta_0) \leq x_k(\delta)$ (Lemma \ref{lemma:inf_prob_convexity} and \eqref{eq:infec_prob}). Thus, it suffices to show that $x_k(\delta_0) > z$. It is easy to see that $\frac{i^2}{1+ic_0z}$ is convex in $i$. By Jensen's inequality, \begin{equation}\label{eq:restrict_c_jensen}
\sum_{i \in \mathcal{D}} \frac{i^2P(i)}{1+ic_0z} \geq \frac{{\langle d\rangle}^2}{1+{\langle d\rangle}c_0z}.
\end{equation}

Since $c_0 > \frac{1}{(1-z)}$ and $\langle d\rangle > 1$, we have
\begin{align*}
&  {\langle d\rangle} < c_0(1-z){\langle d\rangle}^2
\\ \implies & {\langle d\rangle} + {\langle d\rangle}^2 c_0z < c_0 {\langle d\rangle}^2
\\ \implies & 1 < \frac{c_0}{\langle d\rangle} \frac{{\langle d\rangle}^2}{1+{\langle d\rangle}c_0z} \leq \frac{c_0}{\langle d\rangle} \sum_{i \in \mathcal{D}} \frac{i^2P(i)}{1+ic_0z} \text{\quad (from \eqref{eq:restrict_c_jensen})}
\\ \implies & 1 < \sum_{i \in \mathcal{D}} \frac{c_0iq_i}{1+ic_0z}.
\end{align*}
Accordingly, we have $v(\delta_0) > z$ where $v(\delta_0)$ satisfies \eqref{eq:v_consistency}. Furthermore, 
\begin{align*}
& c_0v(\delta_0) > c_0z > \frac{z}{1-z}
\\ \implies & z < \frac{c_0v(\delta_0)}{1+c_0v(\delta_0)} \leq \frac{c_0kv(\delta_0)}{1+c_0kv(\delta_0)} = x_k(\delta_0), 
\end{align*}
for $k \in \mathcal{D}$. This concludes the proof. 
\qed \end{pf}

We are now ready to prove the existence of a DBE.

\begin{proposition}\label{proposition:NE_existence_behavioral}
Let the set of nodes with degree $k$ have weighting function $w_k(\cdot)$ satisfying Assumption \ref{assumption:weightingfunction}. Let $\mathbf{x}_{\min,w_k} = \mathbf{x}_{\min}$, and $c_k = c_0 > \frac{1}{1-\mathbf{x}_{\min}}$ for every $k \in \mathcal{D}$. Then there exists a DBE of the game $\Gamma(\mathcal{D},P,\{c_k\}_{k \in \mathcal{D}},\{w_k\}_{k \in \mathcal{D}})$. 
\end{proposition}
\begin{pf}
From Assumption \ref{assumption:weightingfunction}, we know that $w_k(x_k(\delta_k,\delta_{-k}))$ is convex in $x_k(\delta_k,\delta_{-k})$ for $x_k(\delta_k,\delta_{-k}) \in [\mathbf{x}_{\min},1]$. Furthermore, from Lemma \ref{lemma:inf_prob_convexity}, we know that $x_k(\delta_k,\delta_{-k})$ is convex in $\delta_k$ for a given feasible curing rate vector $\delta_{-k}$. For $\delta \in \prod_{k \in \mathcal{D}} \Delta_k$, $x_k(\delta) > \mathbf{x}_{\min}$ following Proposition \ref{proposition:restrict_c}, and accordingly, $w_k(x_k(\delta_k,\delta_{-k}))$ is convex in $\delta_k$ for a given $\delta_{-k}$. 
From the above discussion, and following the proof of Proposition \ref{proposition:NEexistence_riskneutral}, we observe that $\Gamma$ is equivalent to a strategic game where all nodes of a given degree are controlled by a single player who minimizes a convex cost function. Following \citep{rosen1965existence}, a PNE exists in the equivalent game. Consequently, $\Gamma$ possesses a DBE.
\qed \end{pf}

\begin{remark}
For Prelec weighting functions (i.e., when $w_k(\cdot)$, $k \in \mathcal{D}$ are given by \eqref{eq:prelec}), $\mathbf{x}_{\min,w_k} = \frac{1}{e}$ is independent of $\alpha$. Thus, the above result holds when nodes of different degrees with Prelec weighting functions are heterogeneous vis-a-vis their weighting parameters.
\end{remark}

At the DBE for true expectation minimizers, we showed that the equilibrium curing rates are $0$ when curing costs are larger than $1$ (Proposition \ref{proposition:NEproperties_riskneutral}). In contrast, the following result shows that under nonlinear probability weighting, the curing rates are strictly positive at the DBE (including when the cost parameters are larger than $1$). 

\begin{proposition}\label{proposition:NE_existence_behavioral}
Let $\delta^{\mathtt{NE}}$ be a DBE strategy profile. Then $\delta^{\mathtt{NE}}_k > 0$ irrespective of the curing rate cost $c_k$. 
\end{proposition}
\begin{pf}
We compute the derivative of the cost function $J_k(\delta_k,\delta_{-k})$ in \eqref{eq:cost_prelec} at $\delta_k = 0$ as
\begin{align*}
\frac{\partial J_k}{\partial \delta_k} & = c_k + w'(x_k(\delta_k,\delta_{-k})) \frac{\partial x_k}{\partial \delta_k}\bigg\rvert_{\delta_k = 0} = c_k - w'(1) \frac{1}{kv} < 0,
\end{align*}
since $w'(1-\epsilon) \to \infty$ as $\epsilon \to 0$, following Assumption \ref{assumption:weightingfunction}. Therefore, the expected perceived cost is decreasing at $\delta_k = 0$, and accordingly $\delta^{\mathtt{NE}}_k > 0$.
\qed \end{pf}

{\bf Discussion}: The above result shows that players with nonlinear perception of probabilities always choose a nonzero curing rate at equilibrium irrespective of the per-unit cost of curing rate (as long as the cost is finite), in contrast with the equilibria under true expectation minimizers. This is a consequence of underweighting of large probabilities. When the true infection probability is $1$, a small increase in curing rate leads to a large perceived reduction in infection probability which leads to a nonzero curing rate at the equilibrium. 

We now illustrate how the nonzero curing rate varies with the per-unit cost in the more tractable case of degree-regular networks.

\subsection{Comparison of curing rates in degree-regular graphs}\label{sub:dreg}

A network is degree-regular when every node has an identical degree $d$. Accordingly, in our framework, an identical curing rate $\delta \geq 0$ is chosen for all nodes in the network. Since the network is degree-regular, a randomly chosen neighbor also has degree $d$. Therefore, $v = x_d$. From \eqref{eq:v_consistency}, we obtain 
$$1 = \frac{d}{\delta+dv} \implies v = 1 - \frac{\delta}{d}.$$
Therefore, the infection probability of a node in the endemic state is
\begin{align}\label{eq:x_d_regular}
x_d = \begin{cases} 
      \hfill 1-\frac{\delta}{d} \hfill & \text{if $\delta \leq d,$} \\
      \hfill 0 \hfill & \text{otherwise}. \\
\end{cases} 
\end{align}

\begin{figure}[t]
	\centerline{\includegraphics[scale = 0.65]{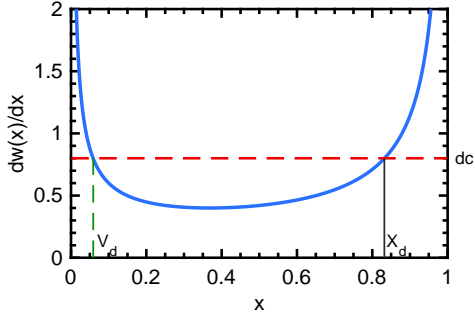}}
	\caption{Roots of $w'(x) = dc$ are denoted by $V_d$ and $X_d$. In this example $w(\cdot)$ is a Prelec weighting function with parameter $\alpha=0.4$ and $dc = 0.8$.}
	\label{fig:prelecweightingfoc}
\end{figure}

Note that for degree-regular graphs, the infection probabilities at the endemic state under DBMF and NIMFA coincide. We focus on the regime where the curing cost $c > \frac{1}{d}$. Let $w(\cdot)$ (satisfying Assumption \ref{assumption:weightingfunction}) be the probability weighting function of the decision-maker. We denote the optimal curing rate for a true expectation minimizer, and under nonlinear probability weighting by $\delta^{\mathtt{N}}$ and $\delta^{\mathtt{W}}$, respectively. As shown in \citep{hota2016interdependent} for weighting functions that satisfy Assumption \ref{assumption:weightingfunction}, there are at most two roots of the equation $w'(x) = dc$ for $x \in [0,1]$ denoted by $X_d > \mathbf{x}_{\min,w}$ and $V_d < \mathbf{x}_{\min,w}$ (as depicted in Figure \ref{fig:prelecweightingfoc}). Recall that $\mathbf{x}_{\min,w} := \argmin_{x \in [0,1]} w'(x)$. We obtain the following result on the optimal curing rates denoted by $\delta^{\mathtt{N}}$ and $\delta^{\mathtt{W}}$ for true and nonlinear perception of probabilities, respectively. 

\begin{proposition}
Let $c > \frac{1}{d}$ be the per-unit cost of curing rate. Then, $\delta^{\mathtt{N}} = 0$, while $\delta^{\mathtt{W}} = \min\{\frac{1}{c},d(1-X_d)\}$.
\end{proposition}
\begin{pf}
For $\delta \leq \frac{1}{c} < d$, the expected cost of a true expectation minimizer is $J(\delta) = 1 - \frac{\delta}{d} + c \delta$, which is strictly increasing in $\delta$. Therefore, $\delta^{\mathtt{N}} = 0$.  

On the other hand, the marginal cost under probability weighting is given by $J'^{(w)}(\delta) = w'\left(1 - \frac{\delta}{d}\right)\frac{-1}{d} + c$. For $\delta \in [0,\frac{1}{c}]$, the true infection probability $1 - \frac{\delta}{d} \in [1-\frac{1}{dc},1]$. If $1-\frac{1}{dc} > X_d$, then  $J'^{(w)}(\delta) < 0$ for every $\delta \in [0,\frac{1}{c}]$, and therefore, $\delta^{\mathtt{W}} = \frac{1}{c}$. 

Otherwise, if $V_d < 1-\frac{1}{dc} \leq X_d$, $\delta = d(1-X_d) \in [0,\frac{1}{c}]$ is the only curing rate that satisfies the first order necessary condition of optimality. Since $X_d > \mathbf{x}_{\min,w}$, we also have $w''(X_d) > 0$. Accordingly, $\delta^{\mathtt{W}} = d(1-X_d)$, and the resulting true infection probability is $X_d$. 

Now suppose that $1-\frac{1}{dc} < V_d$. In this case, both $d(1-X_d)$ and $d(1-V_d)$ satisfy the first order optimality condition. First we show that $J^{(w)}(\frac{1}{c}) \leq J^{(w)}(d(1-V_d))$.\footnote{The following arguments are analogous to the ones used in the proof of Lemma 1 in our prior work \citep{hota2016interdependent}.} Let $Z_d := 1-\frac{1}{cd}$. From \eqref{eq:cost_prelec}, we obtain
\begin{align*}
J^{(w)}\left(\frac{1}{c}\right) & = w(Z_d) + 1, 
\\ J^{(w)}(d(1-V_d)) & = w(V_d) + cd(1-V_d).
\end{align*}
Accordingly, 
\begin{align*}
& J^{(w)}\left(\frac{1}{c}\right) - J^{(w)}(d(1-V_d)) 
\\ = & w(Z_d) - w(V_d) + 1 - cd(1-V_d)
\\ = & w(Z_d) - w(V_d) - cd(Z_d-V_d)
\\ = & (Z_d-V_d) \left[\frac{w(Z_d) - w(V_d)}{Z-V_d} - w'(V_d) \right] < 0,
\end{align*}
where the inequality follows from the concavity of $w(x)$ for $x \in [Z_d,V_d]$. On the other hand, $1 = J^{(w)}(0) < J^{(w)}(\frac{1}{c}) = 1+w(1-\frac{1}{cd})$. Finally, $J'^{(w)}(\delta) \leq 0$ for $\delta \in [0,d(1-X_d)]$ with $J'^{(w)}(d(1-X_d)) = 0$ and $J''^{(w)}(d(1-X_d)) > 0$. Thus, $J^{(w)}(d(1-X_d)) \leq J^{(w)}(0)$. Therefore, $\delta^{\mathtt{W}} = d(1-X_d)$ in this case as well.
\qed \end{pf}

In other words, when the per-unit cost of curing satisfies $c > \frac{1}{d}$, the optimal curing rate for a true expectation minimizer is $0$, and consequently the infection probability is $1$. In contrast, a decision-maker with nonlinear perception of probabilities chooses a nonzero curing rate which decreases to $0$ in a smooth manner as $c$ increases. Even for a large per-unit cost of curing rate, the infection probability is less than $1$ under nonlinear probability weighting.


\section{Conclusion}
In this paper, we initiated the study of strategic decision-making by humans to protect against SIS epidemics on networks. We considered a population game framework where nodes choose curing rates to reduce the infection probability in the endemic state of the SIS epidemic under suitable mean-field approximations. We established the existence of degree based equilibria in both settings under risk neutral as well as behavioral decision-makers whose perceptions of infection probabilities are governed by prospect theory. Furthermore, we showed that players with nonlinear perception of infection probabilities always choose a nonzero curing rate at the equilibrium, while true expectation minimizers choose the curing rate to be zero for sufficiently high cost per-unit cost of curing. Characterizing the price of anarchy as well as the social costs at the equilibria under true and nonlinear probability weighting remain as important future directions.

\bibliography{refs}             

\end{document}